\makeatletter
\declare@file@substitution{revtex4-1.cls}{revtex4-2.cls}
\makeatother
\documentclass[twocolumn]{aastex631}  
\usepackage{graphicx}
\usepackage{natbib}
\usepackage{graphics}
\usepackage{amsmath}
\usepackage{amssymb}
\usepackage{gensymb}
\usepackage{comment}
\usepackage{hyperref}
\usepackage{multirow}
\usepackage{xcolor}
\usepackage[page]{appendix}

\begin{document}
\title{Influence of Modeling Assumptions on the Inferred Dynamical State of Resonant Systems: A Case Study of the HD 45364 System}
\published{2025 February 19}
\author[0009-0005-9428-9590]{Ian Chow}
\affiliation{David A. Dunlap Department of Astronomy \& Astrophysics, University of Toronto, 50 St George Str, Toronto, ON M5S 3H4, Canada}
\affiliation{Department of Physics and Astronomy, University of Western Ontario, 1151 Richmond Str, London, ON N6A 3K7, Canada}
\affiliation{Institute for Earth and Space Exploration, University of Western Ontario, Perth Drive, London, ON N6A 5B7, Canada}

\author[0000-0002-1032-0783]{Sam Hadden}
\affiliation{Canadian Institute for Theoretical Astrophysics, 60 St George Str, Toronto, ON M5S 3H8, Canada}
\begin{abstract}
Planetary systems exhibiting mean-motion resonances (MMRs) offer unique opportunities to study the imprint of disk-induced migration on the orbital architectures of planetary systems.
The HD 45364 system, discovered via the radial velocity (RV) method to host two giant planets in a 3:2 MMR, has been the subject of several studies attempting to reconstruct the system's orbital migration history based on its present-day resonant configuration.
Recently, \citet{li_new_2022} called into question the system's residence in the 3:2 MMR based on a revised orbital solution derived from an expanded set of RV observations that extend the time baseline of the original discovery data by over a decade.
However, we show that inferences about the planets' dynamical state with respect to the 3:2 MMR are sensitive to the particular prior assumptions adopted in the orbital modeling.
Using $N$-body dynamical models, we show that orbital solutions constrained to reside deep in the 3:2 MMR fit the RV data with a similar quality to unconstrained orbital solutions.
We conclude that the RV observations of HD 45364 are consistent with orbital configurations produced by smooth migration and resonance capture. We further show that past convergent orbital migration can reproduce the system's present-day orbital configuration provided that the ratio of migration to eccentricity damping timescales, $K$, was in the range $11\lesssim K\lesssim 144$. We also find that dynamical interactions in the system can break the usual mass-inclination degeneracy inherent to Keplerian models of RV observations and constrain the planets' absolute masses to within a factor of $\sim1.5$.
\end{abstract}


\section{Introduction}
Numerous pairs of giant planets in mean-motion resonances (MMRs) have been discovered via the radial velocity (RV) method \citep[e.g.][]{mayor_coralie_2004, lee_21_2006, tinney_21_2006, correia_harps_2009, niedzielski_substellar-mass_2009, johnson_retired_2010, johnson_retired_2011, wright_california_2011, robertson_second_2012, wittenmyer_detailed_2013, wittenmyer_pan-pacific_2016, giguere_newly_2015, luque_precise_2019, trifonov_two_2019}. The resonant orbital configurations of these systems are generally interpreted as products of past convergent orbital migration induced by planet interactions with the protoplanetary disk \citep{goldreich_disk-satellite_1980}, as such migration generically causes planets to be captured into any low-order MMRs they encounter. 
Thus, systems hosting resonant planets present attractive choices for studying orbital migration and its role in planetary formation.

One such system is HD 45364, which consists of a K0V star with mass of $0.82~{M}_\odot$ and two approximately Jupiter-mass planets with orbits near a 3:2 commensurability, discovered by \citet{correia_harps_2009} using the High Accuracy Radial Velocity Planet Searcher \citep[HARPS;][]{mayor_setting_2003}.
The original best-fit RV solution of \citet{correia_harps_2009} indicated that the two planets were on slightly eccentric orbits and experienced large librations within the $3$:$2$ MMR, with $\sim  70\degree$ excursions of the resonant angles away from their equilibrium values. Subsequent work by \cite{rein_dynamical_2010} used hydrodynamical simulations of planet-disk interactions to find potential migration histories that could produce orbital configurations that were consistent with the observed RV data. They found orbital solutions arising from their migration simulations that differed significantly from the best-fit solution reported by \citet{correia_harps_2009} but which reproduced the observed data with approximately equal statistical significance. Similar conclusions were reached by \citet{correa-otto_new_2013} using $N$-body simulations with parameterized migration and eccentricity damping forces.
Finally, \citet{hadden_modeling_2020} showed, using a Bayesian model comparison framework, that the available RV data for HD 45364 were fully consistent with the planet pair residing in a zero-libration amplitude resonant configuration that would naturally arise under the effects of smooth migration and eccentricity damping forces.   

New RV measurements of HD 45364 by the HARPS team, available through the HARPS-RVBank archive \citep{trifonov_public_2020}, as well as observations from the High Resolution Echelle Spectrometer \citep[HIRES;][]{vogt_hires_1994} instrument, obtained by \citet{li_new_2022}, have significantly extended the observational baseline of the original data set analyzed in the studies described above.
Based on orbital fitting of this extended data set, \citet{li_new_2022} conclude that the HD 45364 planets are slightly outside of the 3:2 MMR, contrary to the conclusions of previous studies and the theoretical expectations for any planet pair migrated into a compact configuration via planet-disk interactions.  

However, as emphasized by both \citet{hadden_modeling_2020} and \citet{jensen_inferred_2022}, inferences about the dynamical state of (near-)resonant planet pairs can be particularly sensitive to modeling assumptions. Indeed, \citet{jensen_inferred_2022} show that measurement noise in RV observations tends to produce a bias towards larger libration amplitudes of planets in MMR. This is essentially because orbital configurations with large libration amplitudes occupy significantly more prior volume than low-libration amplitude configurations. 

In this paper, we re-analyze the recently-extended data set of RV measurements for the HD 45634 system in order to explore how the system's inferred dynamical state with respect to the 3:2 MMR is influenced by modeling assumptions. Following \citet{jensen_inferred_2022}, we explore how the system's dynamical state is influenced by the assumed Bayesian priors on planets' orbital parameters. We also show that un-modeled measurement noise can masquerade as a significant libration amplitude in the system. 


This paper is structured as follows: 
In Section \ref{sec:rv_fitting}, we present our orbital fitting procedures,
fit an RV curve to observations with an $N$-body dynamical model and 
estimate posterior distributions for the model parameters.
We then investigate how the system's inferred dynamical state is affected by assumptions about measurement noise and by different priors on the model parameters.
In Section \ref{sec:migration_history} we 
give a brief review of the dynamics of resonance capture and
analyse possible scenarios for the migration of HD 45364's planets assuming an MMR configuration. We summarize our conclusions in Section \ref{sec:discussion_conclusions}.

\section{Radial Velocity Fitting} \label{sec:rv_fitting}

The original RV observations of HD 45364 used by \citet{correia_harps_2009} were taken by the HARPS telescope over a period of approximately $1600$ days from December $2003$ to April $2008$. Further observations of the system by HARPS have since extended its observational baseline to September $2017$, spanning over $5000$ days in total. \citet{trifonov_public_2020} improved the HARPS RV precision by recomputing the RVs from observed spectra using the Spectrum Radial Velocity Analyser pipeline of \citet{zechmeister_spectrum_2018} and correcting a number of systematic errors. The new HARPS data were later published as part of the HARPS-RVBank archive\footnote{\href{https://github.com/3fon3fonov/HARPS\_RVBank}{github.com/3fon3fonov/HARPS\_RVBank}}\citep{trifonov_public_2020}, totalling $122$ RV observations. In addition, an upgrade to HARPS's optical fibres in May $2015$ \citep{lo_curto_harps_2015} changed the instrumental profile and thus the RV offset between the pre- and post-upgrade RV observations \citep{trifonov_public_2020}. As such, we treat the pre- and post-upgrade RV data as taken from two separate instruments in this work. Finally, we also use $7$ RV observations taken by HIRES from December $2009$ to September $2021$, provided in Table $1$ of \citet{li_new_2022}. 

Here we report our fits to the full set of RV data available for HD 45634, consisting of $129$ data points 
spanning a total observational baseline of $\sim6500$ days.\footnote{We note that \citet{li_new_2022} use only $121$ total data points ($114$ from HARPS and $7$ from HIRES) in their analysis, as in two separate cases, they combine five HARPS observations taken over a short time span into a single observation.
}
We begin in Section \ref{sec:baseline_model} by describing our $N$-body dynamical model, which we fit under relatively uninformative priors on the planets' orbital parameters. 
In Section \ref{sec:libration_amplitude}, 
we explore how the inferred dynamical state of the system with respect to the 3:2 MMR is influenced by the model priors and assumptions about the level of measurement noise. 
\subsection[N-body dynamical model]{$N$-body dynamical model}
\label{sec:baseline_model}
We fit the RV observations of HD 45364 using an $N$-body model that accounts for Newtonian gravitational interactions among the star and two planets in order to compute the star's RV.
Integrations for this dynamical model and all other simulations presented in this paper are conducted with the \texttt{REBOUND} \citep{rein_rebound_2012} code for numerical $N$-body integration, using the \texttt{IAS15} integrator \citep{rein_ias15_2015}
for the integration and Jacobi elements for the coordinate frame. 
We fit our model to three sets of RV observations: the pre- and post-upgrade HARPS observations (which we denote HARPS1 and HARPS2 respectively), and the HIRES observations.

The planets' orbits are parameterized in terms of their orbital periods, $P_i$, eccentricities, $e_i$, times of conjunction, $T_i$, and arguments of pericenter, $\omega_i$, all specified at a reference epoch of $t_0 = 54422.79 + 2.4\times10^6$ BJD.
This value is chosen as the median time of the RV observations for the system.
We parameterize the planets' masses, $m_i$, in our model as functions of the planets' orbital parameters and RV semi-amplitudes, $K_i$, according to 
\begin{equation}
    m_i = \left(\frac{2\pi G}{P_i}\right)^{-1/3}M_*^{2/3}\,\frac{K_i}{\sin I}\,\sqrt{1 - e_i^2} \label{eqn:semiamp_to_mass}
\end{equation}
where $M_*$ is the host star mass (which we fix to be $M_* = 0.82\,M_\odot$), and $G$ is Newton's universal gravitational constant. We assume the planets' orbits to be coplanar but allow their orbital planes to be inclined with respect to the sky plane by an angle, $I$, and include $\sin(I)$ as a free parameter in our $N$-body model. Our model also includes RV offsets, $\gamma_j$, for the three sets of observations HARPS1, HARPS2, and HIRES.
Finally, our model includes three 
instrumental jitter parameters $\sigma_{\mathrm{jit}, j}$, one for each set of observations. Writing $\pmb{\theta}_\mathrm{orb} = \{\sin I\} \cup \{K_i,P_i,e_i,T_i,\omega_i\}_{i=1}^{2}$ to denote the subset model parameters governing the orbital data, the log-likelihood of a set of model parameters can be written as 

\begin{align}
   \ln\mathcal{L}\left(\theta,\gamma_j, \sigma_{\mathrm{jit}, j}\right)= &-\sum_{j=1}^{3}\sum_{i=1}^{N_j}\,\frac{\left(v_{j,i} - \gamma_j -  \bar{v}_{j,i}(\pmb{\theta}_\mathrm{orb})\right)^2}{2\left(\sigma_{j,i}^2 + \sigma_{\mathrm{jit}, j}^2\right)} \nonumber \\
   &- \sum_{i=1}^{N
   _j}
   \,\ln\sqrt{2\pi\left(\sigma_{j, i}^2 + \sigma_{\mathrm{jit}, j}
   ^2
   \right)} 
   \label{eqn:log_likelihood}
\end{align}
where the index $j$ runs over the three sets of RV data from HARPS1, HARPS2, and HIRES, $v_{j,i}$ denotes $i$th observed RV from the $j$th set of observations, $\sigma_{j,i}$ denotes this velocity's reported measurement uncertainty, and $\bar{v}_{j,i}(\pmb{\theta}_\mathrm{orb})$ denotes the star's RV predicted by our $N$-body dynamical model at the time of observation.
%

We use Markov Chain Monte Carlo (MCMC) sampling to estimate the posterior distribution of our model parameters. We adopt priors that are uniform on the interval $[0,\infty)$ for planets' orbital periods, $P_i$, semi-amplitudes, $K_i$, and times of conjunction, $T_i$. Our prior on the sine of the planets' orbital inclination, $\sin I$, is uniform between 0 and 1. We adopt the standard procedure of parameterizing planets' orbital eccentricities, $e_i$, and arguments of pericenter, $\omega_i$, by the variables $\sqrt{e_i}\cos\omega_i$ and $\sqrt{e_i}\sin\omega_i$ when sampling, while imposing a prior that planets' eccentricities be less than 1. Finally, our priors on the value of the instrumental jitters, $\sigma_{\mathrm{jit}, j}$, are uniform between 0 and 10 m s$^{-1}$.
The MCMC sampling is carried out using the \texttt{emcee} \citep{goodman_ensemble_2010, foreman-mackey_emcee_2013} software package.
We initialize $50$ chains in a multivariate Gaussian distribution centered on the best-fit parameters obtained by minimizing the log-likelihood. 
The chains are then evolved for $50,000$ steps, generating a total of $2.5$ million posterior samples.
We compute the autocorrelation lengths of our MCMC chains for each model parameter and estimate that our simulations produce $\gtrsim3000$ independent posterior samples by dividing the total number of samples by the longest computed autocorrelation length.

\begin{figure*}[t]
    \centering
    \includegraphics[width=1.05\linewidth]{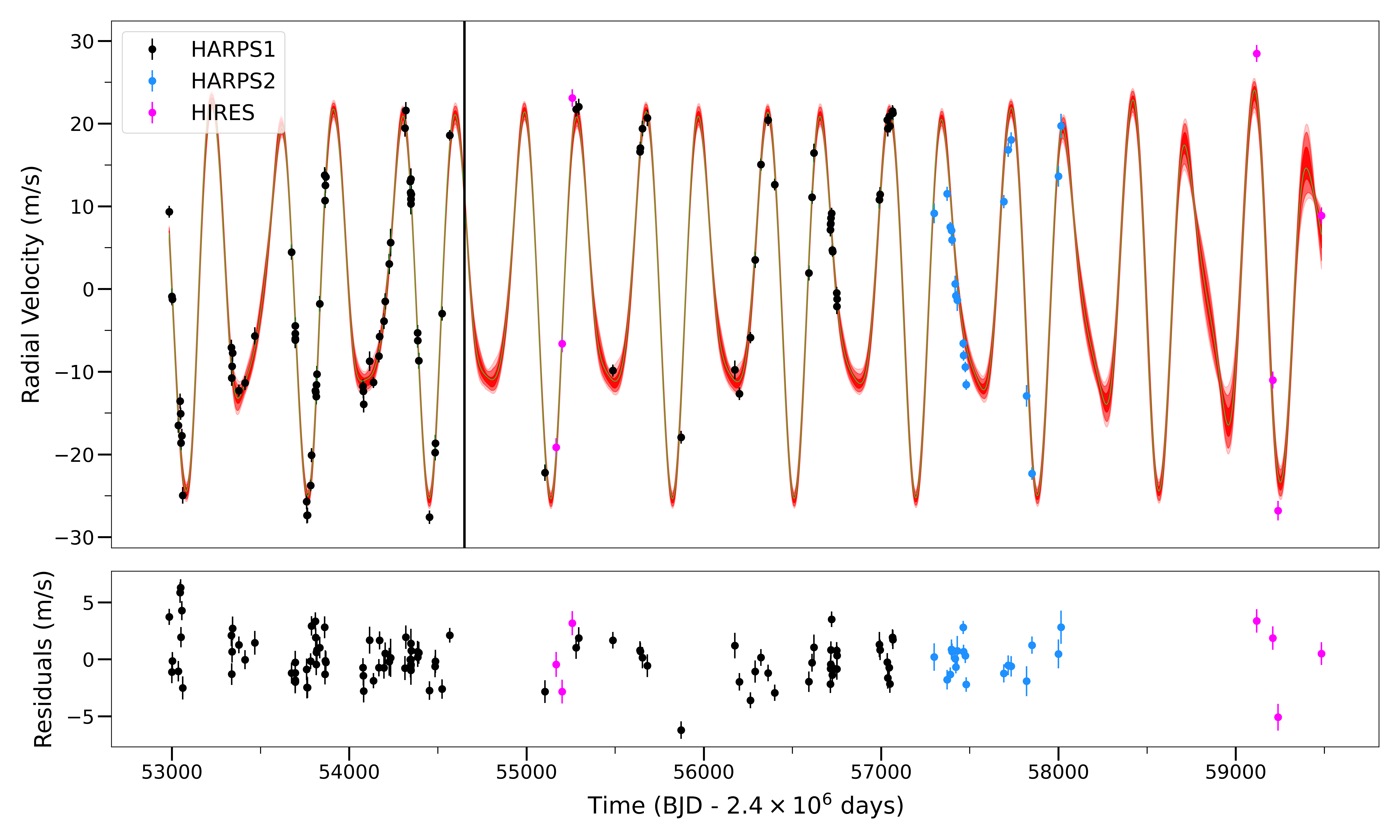}
      \caption{Summary of RV fitting results and uncertainties from the MCMC posterior distributions for HD 45364, using data from the pre- and post-upgrade HARPS (HARPS1 and HARPS2, respectively) and the HIRES instruments. The top panel shows our fit RV signal plotted over the observed RVs, with $1\sigma$, $2\sigma$ and $3\sigma$ uncertainties illustrated by the red  shaded regions.
      The maximum log-likelihood solution is plotted as the green line. 
      The HARPS data points to the right of the black line are new data obtained from the HARPS-RVBank archive \citep{trifonov_public_2020} since the original analysis of \citet{correia_harps_2009}. Time is given in units of  Barycentric Julian Date (BJD) - $2.4 \times 10^6$ days. 
      The normalized residuals of the best-fit solution are plotted in the bottom panel.
      }
     \label{fig:rv_curve}
\end{figure*}

The resulting
 RV fit  is shown  in Figure \ref{fig:rv_curve}.
The top panel shows 
the $1\sigma$, $2\sigma$ and $3\sigma$ uncertainties of our fit RV curve 
together with the observed data. We obtain best-fit instrumental jitter values of $1.41$, $0.694$ and $3.00$ m s$^{-1}$ for the HARPS$1$, HARPS$2$ and HIRES telescopes, respectively.
The maximum log-likelihood solution and the $5\%$, $50\%$ and $95\%$ quantiles of the $1$D posterior distributions obtained from MCMC sampling are reported in Table \ref{tab:fit_params}.

\begin{table*}[t]
    \begin{center}
    \caption{Maximum Log-Likelihood Fit and MCMC Posterior Quantiles for Fitted Parameters
    }
    \begin{tabular}{|c|c|c c c|}
    \hline
    \textbf{Parameter} & \textbf{Maximum Log-Likelihood} & \multicolumn{3}{c|}{\textbf{MCMC Posterior Quantiles}} \\
    \hline
    $\ln\mathcal{L}$ &  $-246.78$ & \multicolumn{3}{c|}{} \\
    \hline
    & & $5\%$ & $50\%$ & $95\%$ \\
    \hline
    $m_b\,\left[M_J\right]$ & $0.191$ & $0.191$ & $0.222$ & $0.320$ \\
    $P_b$ [days] & $227.88$ & $227.20$ & $227.93$ & $228.69$ \\
    $e_b$ & $0.0505$ & $0.0328$ & $0.0613$ & $0.0917$ \\
    $T_{b}$ [BJD - $2.4 \times 10^6$ days] & $52799.70$ & $52793.08$ & $52801.02$ & $52808.58$ \\
    $\omega_b$ [rad] & $-2.95$ & $-3.03$ & $-2.51$ & $2.94$ \\
    \hline
    $m_c\,\left[M_J\right]$ & $0.550$ & $0.557$ & $0.641$ & $0.932$ \\
    $P_c$ [days] & $344.06$ & $343.53$ & $344.01$ & $344.49$ \\
    $e_c$ & $0.00991$ & $0.00132$ & $0.0136$ & $0.0353$ \\
    $T_{c}$ [BJD - $2.4 \times 10^6$ days] & $52985.87$ & $52982.85$ & $52986.16$ & $52989.33$ \\
     $\omega_c$ [rad] & $1.34$ & $-1.99$ & $1.03$ & $2.51$ \\
     \hline
     $\sin\left(I\right)$ & $1.00$ & $0.594$ & $0.860$ & $0.988$ \\
     $\sigma_{\mathrm{jit}, \mathrm{HARPS1}}$ [m/s] & $1.41$ & $1.19$ & $1.50$ & $1.76$ \\
     $\sigma_{\mathrm{jit}, \mathrm{HARPS2}}$ [m/s] & $0.694$ & $0.17$ & $0.96$ & $1.66$ \\
     $\sigma_{\mathrm{jit}, \mathrm{HIRES}}$ [m/s] & $3.00$ & $1.89$ & $4.00$ & $7.57$ \\
     \hline
     \end{tabular}
    \end{center}
    \textbf{Note}. The orbital periods $P_i$, times of conjunction $T_i$, eccentricities $e_i$ and arguments of pericenter $\omega_i$ are osculating parameters valid for the reference epoch $t_0 = 2454422.79$ BJD.
    \label{tab:fit_params}
\end{table*}

\begin{figure*}[t]
\centering
\begin{minipage}{.49\textwidth}
  \centering
  \includegraphics[width=1.\linewidth]{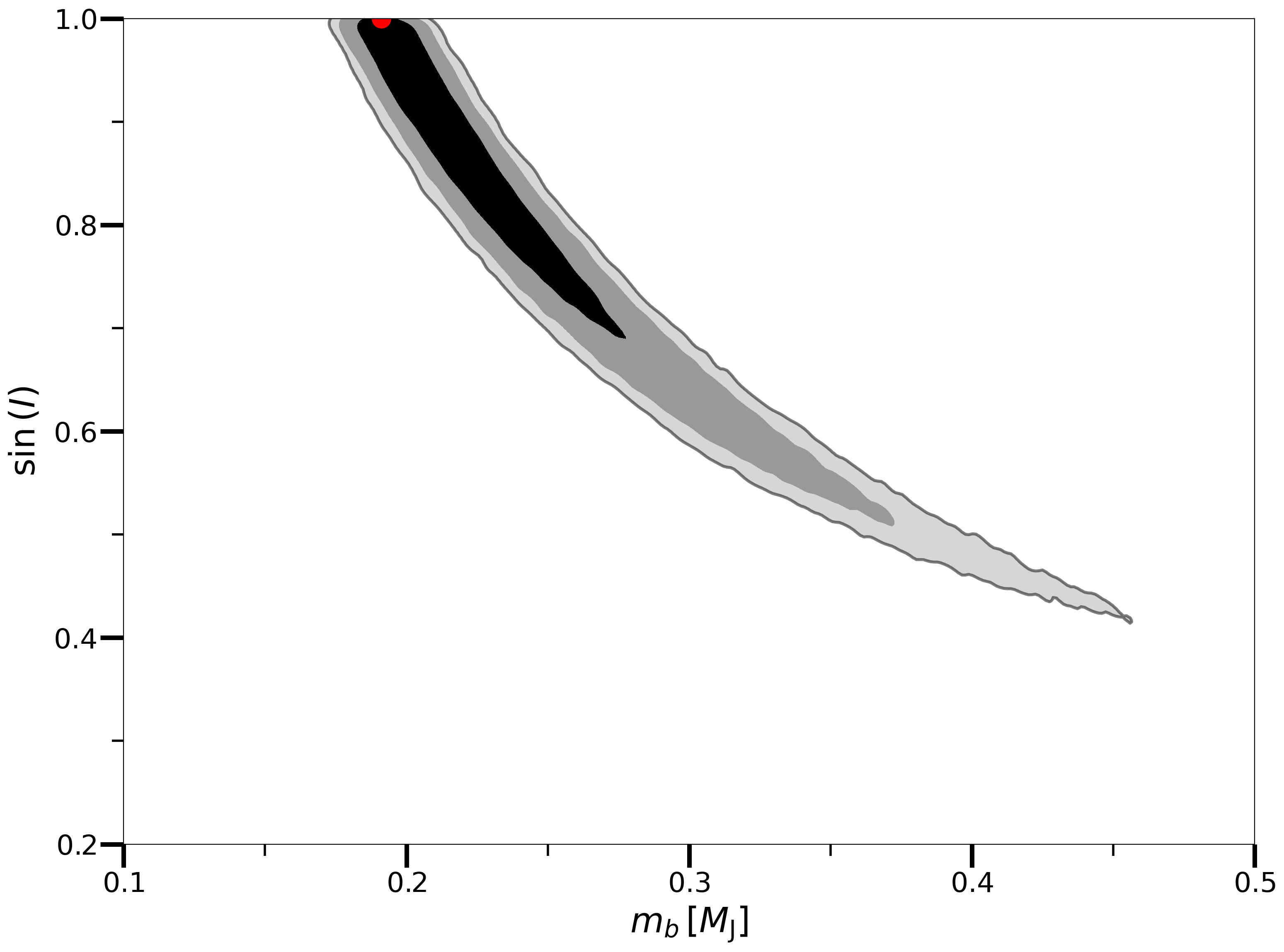}
\end{minipage}
\begin{minipage}{.49\textwidth}
  \centering
  \includegraphics[width=1.\linewidth]{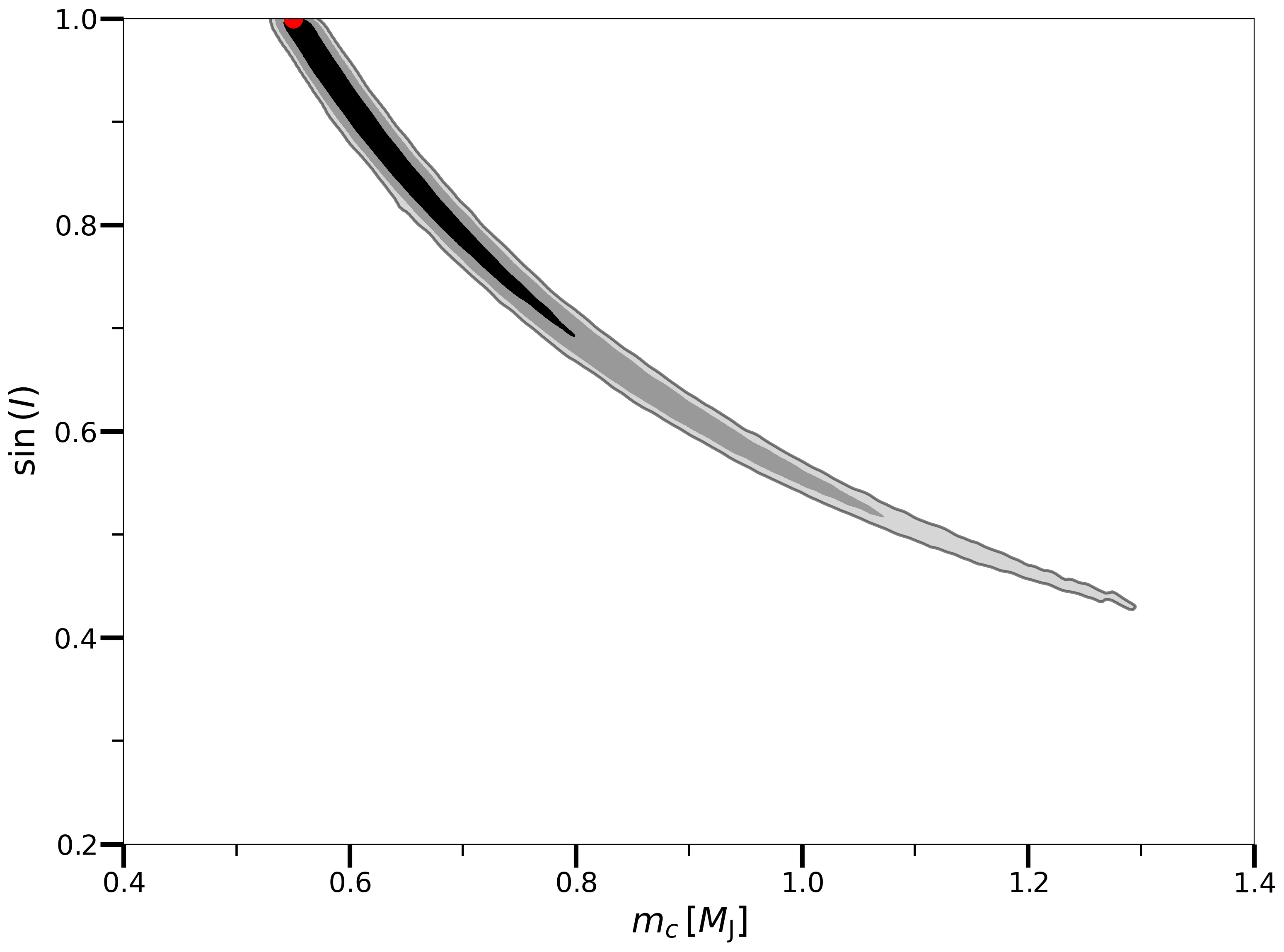}
\end{minipage}
\caption{The marginal $2$D MCMC posterior distributions of $\sin\left(I\right)$  against planet masses $m_b$ (left) and $m_c$ (right), with the planets assumed to be in coplanar orbits. The contours show the $1\sigma$, $2\sigma$ and $3\sigma$ bounds of the posterior distribution. 
The points corresponding to the best-fit solution obtained by maximizing the log-likelihood $\ln \mathcal{L}$ as given in Table \ref{fig:rv_curve} are marked in red.}
\label{fig:mass_sin_i_dist}
\end{figure*}

The large masses and close orbits of the two planets in HD 45364 give rise to mutual gravitational interactions between the planet pair that are strong enough for our dynamical model to break the usual $m\sin I $ degeneracy inherent to purely Keplerian models. 
We therefore let inclination vary as a free parameter in our model, allowing us to place constraints on the planet masses and inclinations.
Using our dynamical model, we constrain the absolute masses of HD 45364 b and c to within a factor of $\sim$$1.5$  of the median values. Figure \ref{fig:mass_sin_i_dist} shows marginalized joint posterior distributions of planet masses versus $\sin I$. The posterior density falls off sharply for 
$\sin I \lesssim 0.6$. 
Furthermore, we find through $N$-body experiments that the dynamical stability of the system demands that $\sin I \gtrsim 0.1$.\footnote{The $N$-body dynamical model MCMC simulations by \citet{li_new_2022} instead assume the planet pairs' inclinations are fixed at $90^\circ$. They do, however, perform maximum-likelihood fits to the RV data for a series of fixed system inclinations. Comparing the maximum likelihoods as a function of system inclination, they find similar results, concluding that $\sin I\gtrsim 0.6$.}

\subsection{The influence of model priors and measurement noise on resonant dynamics} 
\label{sec:libration_amplitude}
\begin{figure*}
    \centering
    \includegraphics[width=1.0\textwidth]{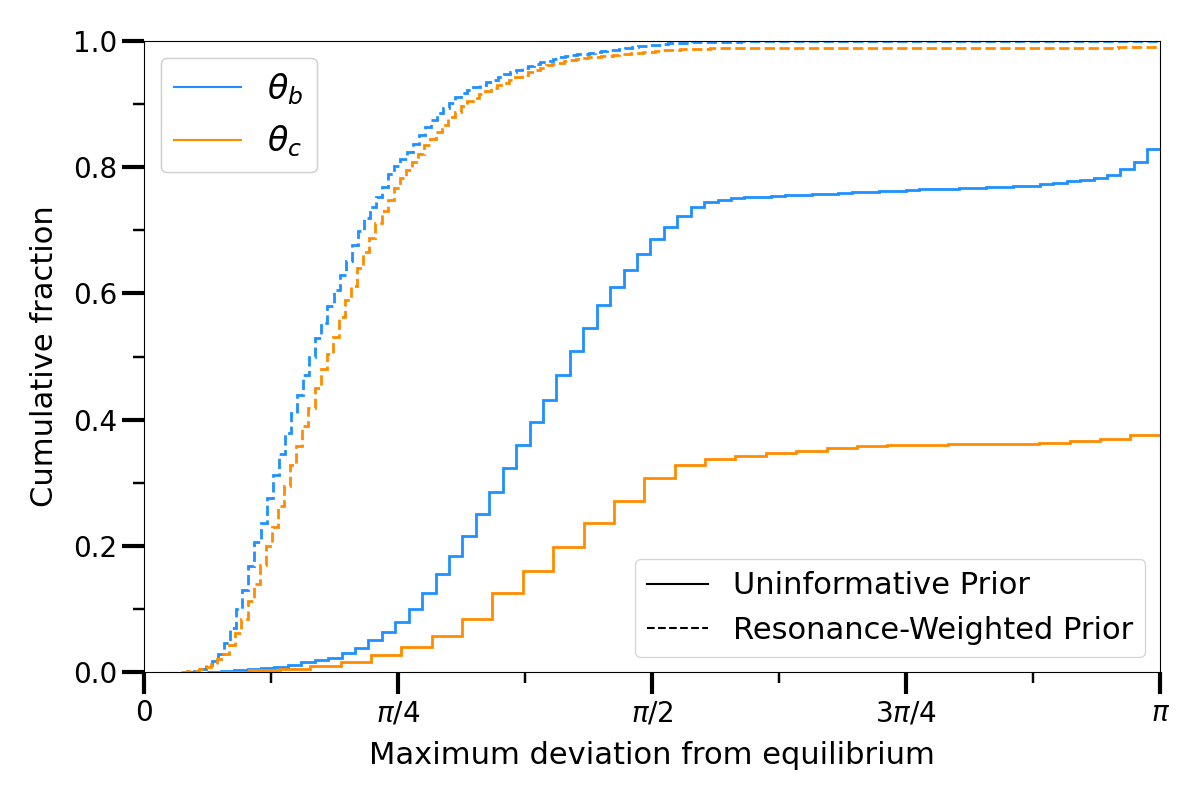}
    \caption{
    The cumulative distribution of the maximum deviation from equilibrium for the resonant angles $\theta_b$ (blue) and $\theta_c$ (orange), measured over short $N$-body integrations, of $50,000$ samples drawn randomly from the MCMC posteriors for the $N$-body dynamical model fit under uninformative priors described in Section \ref{sec:baseline_model} (solid lines) and the model described in Section \ref{sec:libration_amplitude} fit under a resonance-weighted prior (i.e. weighted toward small libration amplitudes), with $S = 0.1$ (dashed lines).
     A large fraction of orbital configurations in the uninformative prior model's MCMC posteriors circulate or have large libration amplitudes ($>90\degree$). In contrast, virtually all resonant angles in the resonance-weighted prior model's MCMC posteriors have small libration amplitudes ($<90\degree$), showing that the chosen model priors strongly affect the inferred libration amplitudes of the system.
    }
    \label{fig:lib_amp_cdfs}
\end{figure*}

The solid lines in Figure \ref{fig:lib_amp_cdfs} show the cumulative distribution for the maximum deviation of the planets' resonant angles from equilibrium, measured from short $N$-body integrations spanning $500$ orbits of the inner planet for a random selection of $50,000$ samples from our full MCMC posterior. For each posterior sample, the resonant angles $\theta_b = 3\lambda_c - 2\lambda_b - \varpi_b$ and $\theta_c = 3\lambda_c - 2\lambda_b - \varpi_c$ were recorded at $5000$ uniformly spaced times in the integration and the maximum deviations from their respective equilibrium values of $0$ and $\pi$ were computed. Figure \ref{fig:lib_amp_cdfs} clearly indicates that most of the posterior samples have large libration amplitudes, with roughly half of the posterior samples showing libration amplitudes greater than $90^\circ$ in both resonant angles and 
a majority of posterior samples ($\gtrsim 60\%$)
showing circulation in angle $\theta_c$. These results nominally support the conclusion of \citet{li_new_2022} that the HD 45364 planets are not fully~locked in the 3:2 MMR.\footnote{Strictly speaking, the circulation of $\theta_c$ does not preclude the system's residence in the 3:2 MMR \citep[see, e.g.][]{petit_resonance_2020}. However, resonance capture via smooth migration is expected to produce resonant configurations with small libration amplitudes, nominally at odds with the posterior distribution of libration amplitudes.}

However, \citet{jensen_inferred_2022} demonstrate that measurement noise can cause a systematic bias to resonant libration amplitudes inferred from RVs. This bias is essentially caused by the fact that higher libration amplitude dynamical configurations occupy a much larger phase-space volume. Consequently, under model priors that are relatively uniform over phase space, noisier data will lead to broader posterior distributions that, in turn, become increasingly skewed towards large libration amplitudes or circulating resonant angles. In this section, we explore the sensitivity of the inferred resonant libration amplitudes of the HD 45364 planets to assumptions about measurement noise and model priors.

\begin{figure*}
    \centering
    \includegraphics[width=1.\textwidth]{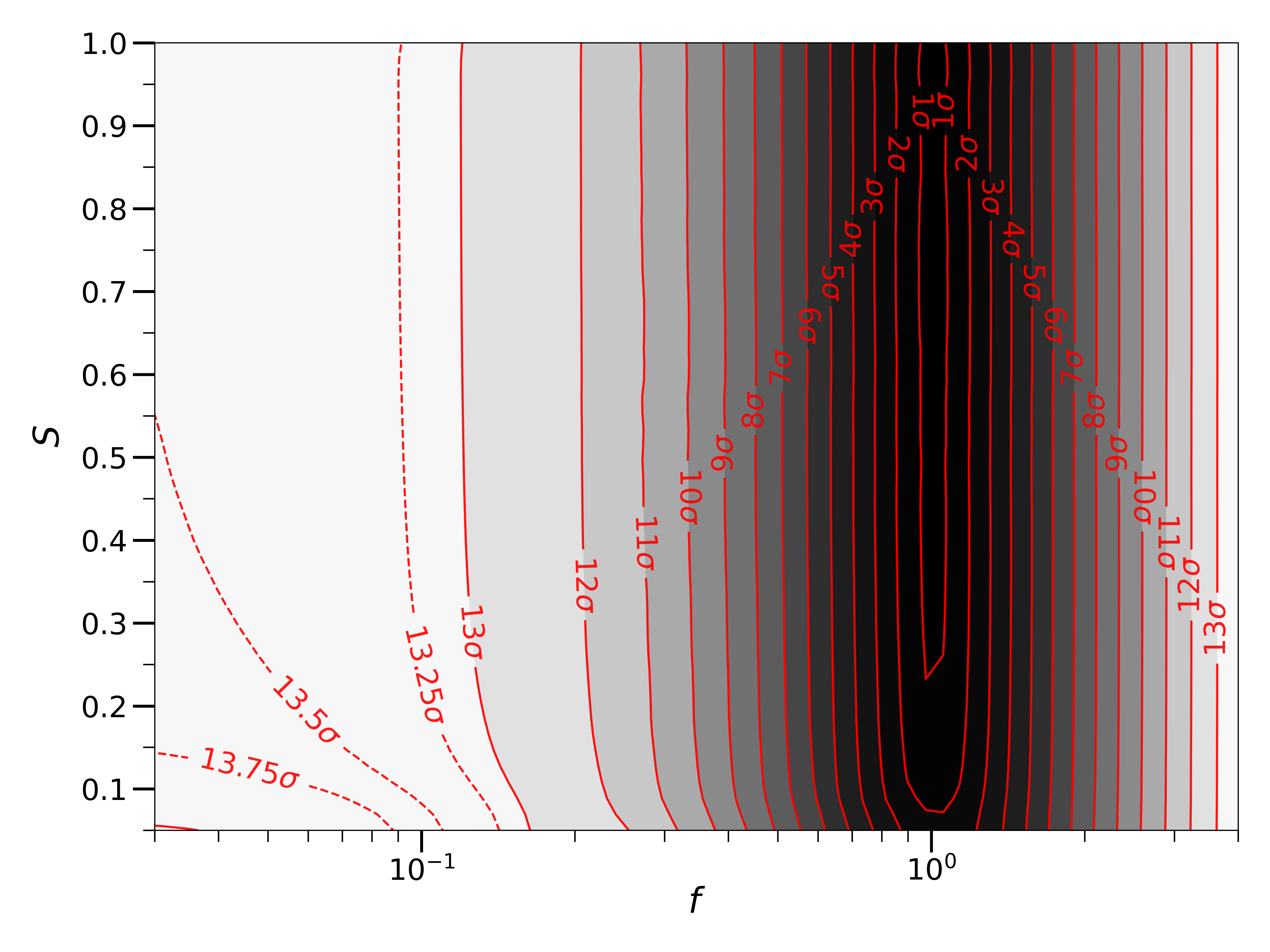}
    \caption{
    A contour plot of the log-likelihood $\ln \mathcal{L}$
    of best-fit orbital solutions found by maximizing the \emph{modified} log-likelihood, $\ln \mathcal{L}'$, given by Equation \eqref{eqn:logl_modified}, for different levels of the
    jitter scale factor, $f$, and 
    libration amplitude penalty, $S$.
    Contours labeled $n\sigma$ mark the log-likelihood levels 
    $\Delta\ln \mathcal{L} = -  \frac{1}{2}n^2$,
    where $\Delta\ln{\mathcal{L}}$ is the log-likelihood function minus its
    global maximum value.
    Dashed contours have been linearly spaced at intervals of $0.25\sigma$ between $13\sigma$  and $14\sigma$  to show that low libration amplitude configurations may be erroneously ruled out compared to high libration amplitude configurations when instrumental jitter is not properly accounted for. 
    }
    \label{fig:Alib_jitter_0_5}
\end{figure*}

To explore the influence of model priors on the inferred resonant behaviour of the HD 45364 planets, we follow \citet{jensen_inferred_2022} and test a series of priors that penalize dynamical configurations according to their libration amplitudes measured in short $N$-body integrations. Specifically, we construct a series of 
priors 
weighted towards resonant configurations
in which we multiply the standard prior probability of a set of model parameters described above in Section \ref{sec:baseline_model} by a term $\propto\exp\left[-\frac{1}{2S^2}A_\mathrm{lib}^2\right]$, where 
$A_\mathrm{lib}$ is the libration amplitude measured as the root-mean-square difference between the resonant angles $\theta_b = 3\lambda_c - 2\lambda_b - \varpi_b$ and $\theta_c = 3\lambda_c - 2\lambda_b - \varpi_c$ and their respective equilibrium values of 0 and $\pi$. For a given set of model parameters, we record the values of resonant angles $\theta_b$ and $\theta_c$ at {$1000$} uniformly spaced times from $N$-body integrations spanning 500 orbits of the inner planet and compute $A_\mathrm{lib}$ according to
\begin{equation}
    A_{\mathrm{lib}}^2 = \frac{1}{{1000}}\sum_{j = 1}^{{1000}}\, \left(\theta_{b_j}^2 + \left(\theta_{c_j} - \pi\right)^2\right)~.
\end{equation}
In Figure \ref{fig:Alib_jitter_0_5} we explore the relationship between the assumed level of measurement noise and the inferred libration amplitudes.
In order to do so, we first construct a modified likelihood function, $\mathcal{L}'$, that depends on the orbital model parameters, $\pmb{\theta}_\mathrm{orb}$, RV offsets, $\gamma_j$, a jitter scale factor, $f$, and a libration amplitude scale factor, $S$, according to 
\begin{equation}
    \ln\mathcal{L}'(\pmb{\theta}_\mathrm{orb}, \gamma_j;f,S) = \ln\mathcal{L}(\pmb{\theta}_\mathrm{orb}, \gamma_j,f\hat{\sigma}_{\mathrm{jit},j})
    - \frac{A_\mathrm{lib}^2(\pmb{\theta}_\mathrm{orb})}{2S^2} \label{eqn:logl_modified}
\end{equation}
where $\mathcal{L}$ is the likelihood function of our uninformative prior RV model given in Equation \eqref{eqn:log_likelihood} and $\hat{\sigma}_{\mathrm{jit},j}$ are the best-fit jitter values determined via a maximum-likelihood fit to the RV data using the uninformative prior RV model. In Figure \ref{fig:Alib_jitter_0_5}, we numerically maximize the modified log-likelihood, $\ln\mathcal{L}'$, on a grid of fixed values for the parameters $f$ and $S$.
We then plot contour levels of $\ln\mathcal{L}$, i.e., the \emph{unmodified, uninformative} RV model log-likelihood, evaluated at the numerically determined parameters that maximize the \emph{modified} likelihood, $\mathcal{L}'$. 
The  $n\sigma$ confidence regions are estimated as $\Delta \ln \mathcal{L} = -  \frac{1}{2}n^2$, where $\Delta \ln \mathcal{L}$ is the difference between the  log-likelihood function and its global maximum value.
If the data strongly favored dynamical configurations in which the resonant angles circulated or possessed large libration amplitudes, we would expect to find that the values of $\ln\mathcal{L}$ decrease significantly for sufficiently small values of $S$.
However, this is not what we see in Figure \ref{fig:Alib_jitter_0_5}. Instead, contours of constant likelihood are nearly independent of the assumed value of $S$ except at the lowest values of the jitter scaling factor, $f$.
Thus, we conclude that the prevalence of dynamical configurations with circulating and large libration amplitude resonant angles found among the posterior samples of our uninformative prior RV model is the result of these dynamical configurations occupying larger prior volume, rather than the data disfavoring low libration amplitude configurations.
Furthermore, the fact that low libration amplitude solutions become disfavored relative to higher amplitude solutions for small values of $f$ indicates that underestimated measurement noise can lead one to improperly infer large libration amplitudes.
%

To further explore the influence of prior assumptions on the inferred dynamical state of the HD 45364 system, we run a second MCMC sampling simulation implementing the modified 
``resonance-weighted"
prior, $\propto\exp\left[-\frac{1}{2S^2}A_\mathrm{lib}^2\right]$, with the value of $S$ set to $S = 0.1$.
The MCMC sampling was carried out with the same procedure described in Section \ref{sec:baseline_model}, but using the modified log-likelihood function $\ln\mathcal{L}'$ with $S = 0.1$ instead of the unmodified log-likelihood function $\ln\mathcal{L}$. As in Section \ref{sec:baseline_model}, $50$ chains were initialized in a multivariate Gaussian distribution centered on the best-fit parameters and each evolved for $50, 000$ steps, generating a total of $2.5$ million posterior samples. The chains successfully converged, with an estimated effective sample size of $\gtrsim4000$. 
We compute the resonant angles' maximum deviation from equilibrium for $50, 000$ random samples from the resonance-weighted prior model's MCMC posterior using the same method as for the uninformative prior model, and plot their cumulative distributions as the dashed lines in Figure \ref{fig:lib_amp_cdfs}. We find that virtually all posterior samples generated with our modified priors have resonant angles that librate with small amplitude ($<90\degree$), demonstrating that our choice of model priors affects the inferred libration amplitudes.

%

\section{Migration History}\label{sec:migration_history}
Here we explore possible planetary migration scenarios resulting in the capture of HD 45364's planets into their current orbital configuration, assuming the system is in MMR.
We first provide a brief overview of the dynamics of resonance capture and describe our 
parameterized
model of planetary migration in Section \ref{sec:resonance_capture}. Section \ref{sec:results} shows our results for possible migration histories of HD 45364 under the modified prior used in Section \ref{sec:libration_amplitude}.

\subsection{Resonance Capture} \label{sec:resonance_capture}
Tidal interactions between planets and a proto-planetary disk can cause their orbits to migrate and their eccentricities to damp \citep{lin_tidal_1979, goldreich_disk-satellite_1980}. Pairs of planets undergoing convergent migration will tend to be captured into any first-order MMRs they encounter, provided their migration rates are sufficiently slow \citep[e.g.,][]{murray_solar_2000}.
After encountering a resonance, the planet pair will migrate together while maintaining a fixed period ratio and experience eccentricity growth until the resonant forcing of their eccentricities is counterbalanced by the damping effects of the disk \citep{snellgrove_disc_2001, lee_dynamics_2002}.
For sufficiently smooth and slow migration forces, the pair of planets will eventually settle into an equilibrium configuration in which their resonant angles and eccentricities remain fixed. 
As discussed in \citet{hadden_modeling_2020}, for a given pair of planet masses and a specific $j$:$j-1$ resonance, these equilibrium configurations constitute a one-parameter family of orbital configurations that can be parameterized by the quantity 
\begin{align}
    \mathcal{D} &= \beta_1\sqrt{\alpha}\frac{e_1^2}{2} + \beta_2\frac{e_2^2}{2} 
    - \frac{\beta_1\beta_2\sqrt{\alpha}}{3\left(j\beta_1\sqrt{\alpha} + (j-1)\beta_2\right)}\Delta~, \label{eqn:amd}
\end{align}
where $e_i$ are the planets' eccentricities, $\beta_i=m_i/(m_1+m_2)$, $\alpha \approx \left(\frac{j-1}{j}\right)^{2/3}$ is the pair's semimajor axis ratio, and $\Delta = \frac{j-1}{j}\frac{P_2}{P_1} - 1$ measures the pair's fractional deviation from exact period ratio commensurability.

If the the migration and eccentricity damping effects of planet-disk interactions are approximated as exponential decays so that 
\begin{align}
    \frac{d}{dt}\ln\left(a_i\right)\bigg|_\mathrm{dis} &= -\tau_{a, i}^{-1}
    \label{eq:dlna_dis}
    \\ 
    \frac{d}{dt}\ln\left(e_i\right)\bigg|_\mathrm{dis} &= -\tau_{e, i}^{-1},
    \label{eq:dlne_dis}
\end{align}
then the ultimate equilibrium configuration into which a planet pair settles will be the one for which 
\begin{align}
     \frac{d}{dt}\mathcal{D}\bigg|_\mathrm{dis} &= 0 \approx - \frac{\beta_1\sqrt\alpha e_1^2}{\tau_{e, 1}} - \frac{\beta_2 e_2^2}{\tau_{e, 2}} \nonumber \\
     & - \frac{\beta_1\beta_2\sqrt{\alpha}}{3\left(j\beta_1\sqrt{\alpha} + (j-1)\beta_2\right)}\frac{3}{2\tau_\alpha}\label{eqn:dissipative_d}~,
\end{align}
where $\tau_\alpha :=(\tau_{a,2}^{-1}-\tau_{a,1}^{-1})^{-1}$.
Therefore, the resonant equilibrium reached by a pair of migrating planets depends on the relative strengths of migration and eccentricity damping forces, frequently parameterized as $K_1 = \tau_\alpha/\tau_{e,1}$ and $K_2 = \tau_\alpha/\tau_{e,2}$. 

Figure \ref{fig:posterior_ecc_track} shows the eccentricities of a pair of planets in the family of 3:2 resonant equilibria. The planets' masses are taken to be the median masses of HD 45364 b and c, as inferred from our MCMC simulations in Section \ref{sec:libration_amplitude}.  The equilibrium configurations are found by running a series of $N$-body simulations with eccentricity and semimajor axis damping forces included. We use the \texttt{reboundx} code's \citep{tamayo_reboundx_2020} \texttt{modify\_orbits\_direct} effect to impose dissipative forces of the form of Equation \eqref{eq:dlna_dis} and \eqref{eq:dlne_dis}. We set the planets' eccentricity damping timescales to $\tau_\mathrm{e,b}=\tau_\mathrm{e,c}=: 2\tau_e = 10^4P_b$. The semimajor axis damping timescales, $\tau_{a,i}$, are chosen so that $K = \tau_\alpha/\tau_e$ ranges from 1 to $10^3$ and 
$\tau_{a, b}/\tau_{a, c} = -\frac{m_b}{m_c}\left(\frac{3}{2}\right)^{2/3}$.\footnote{The latter condition is imposed for numerical convenience, as it ensures that the system remains at approximately the same mean stellocentric distance over the course of the simulations, obviating the need to adjust the simulation time step.}
The planets are initially placed in circular, coplanar orbits with a period ratio of  $P_c/P_b = 1.1 \times \frac{3}{2}$, just outside the 3:2 MMR. The simulations are then integrated until the planets reach their equilibrium eccentricities. Plotting the equilibrium eccentricities of each simulation in Figure \ref{fig:posterior_ecc_track} produces an  eccentricity ``track," along which each simulation evolves after capturing into resonance until it reaches its unique stopping point at which Equation \eqref{eqn:dissipative_d} is satisfied \citep[see][]{hadden_modeling_2020}.

\subsection{Results}\label{sec:results}
\begin{figure*}[t]
   \centering
   \includegraphics[width=1.\textwidth]{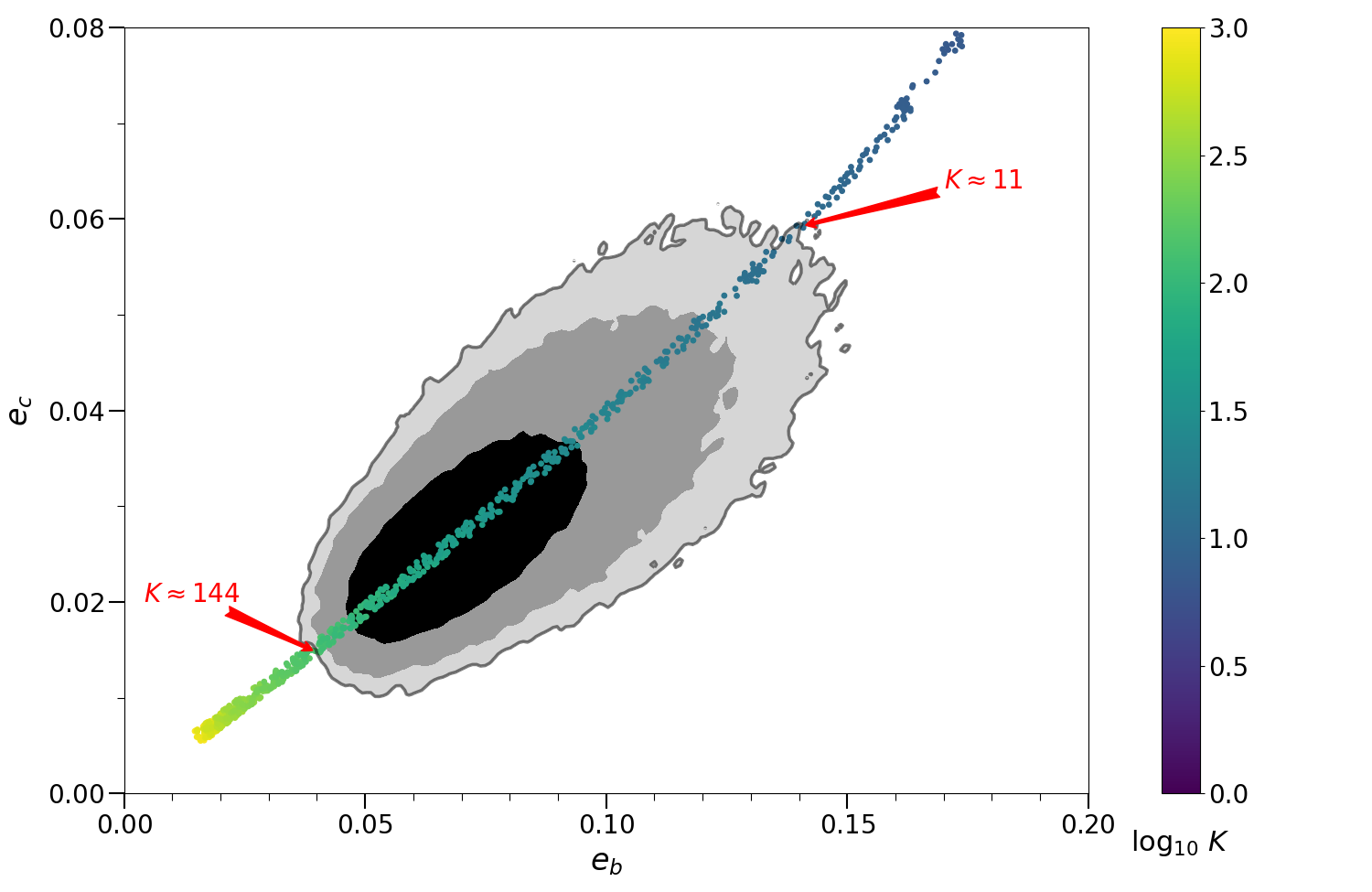}
   \caption{
   Comparison of our resonance capture simulations with the $2$D marginalized distribution of planet eccentricities from the MCMC posterior of the resonance-weighted prior model described in Section \ref{sec:libration_amplitude} with $S = 0.1$.
   The $1\sigma$, $2\sigma$ and $3\sigma$ contours of the posterior distribution are shaded.
   The simulated equilibrium eccentricity track using the best-fit planet masses is overlaid on the posterior distribution, with points falling within the MCMC posterior corresponding to simulated migration scenarios consistent with the observed eccentricities of the system. The approximate $K$ values for the bounds of the overlapping track segment up to the $3\sigma$ contours are labeled as well, suggesting the orbital configuration of HD 45634 b and c could be reproduced with a value 
   $11 \lesssim K \lesssim 144$.
   }
   \label{fig:posterior_ecc_track}
\end{figure*}

We determine possible migration scenarios for HD 45364 based on the posterior distribution of orbital configurations from the resonance-weighted prior model described in Section \ref{sec:libration_amplitude}. Eccentricity posterior samples from our MCMC simulations with this model are plotted in Figure \ref{fig:posterior_ecc_track}. 
Comparing the inferred distribution of the planets' eccentricities to the range of equilibrium eccentricities generated by $N$-body simulations, we can estimate the relative strengths of eccentricity damping to orbital migration that led to the planets' capture into the 3:2 MMR.  The eccentricity posterior samples fall between the equilibria of simulations with $11 \lesssim K \lesssim 144$, indicated by the red arrows in Figure \ref{fig:posterior_ecc_track}.

\begin{figure*}  
   \centering
   \includegraphics[width=1.\textwidth]{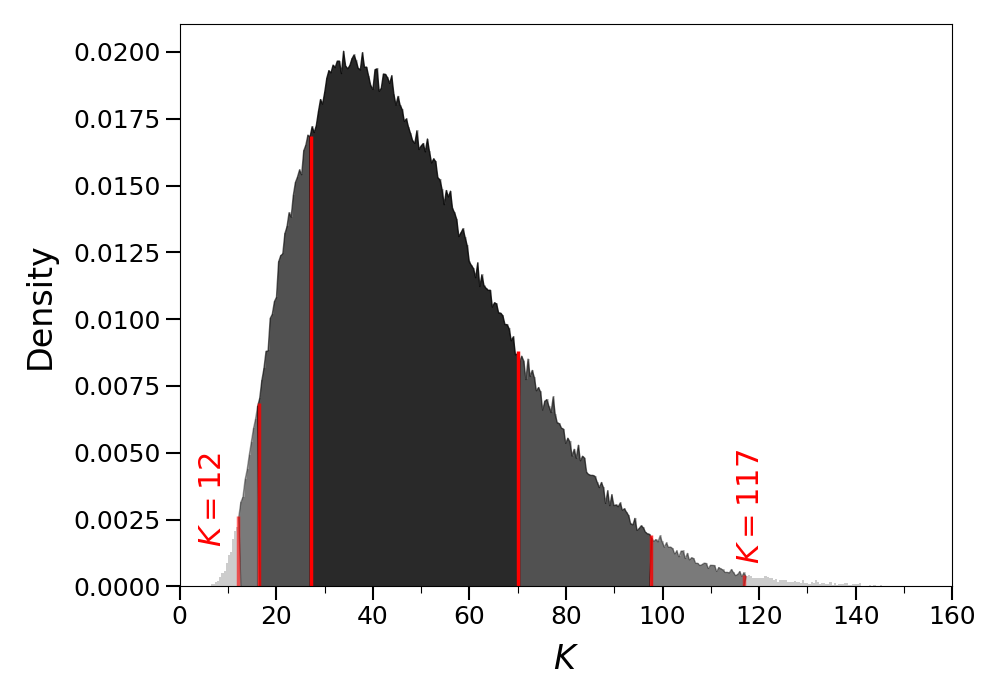}
   \caption{
   Density histogram of $K$ values for the resonance-weighted prior model's posterior eccentricity distribution shown in Figure \ref{fig:posterior_ecc_track},
   with the $68\%$, $95\%$ and $99\%$ credible intervals marked by red lines.
   The $K$ values are computed from the posterior distribution using the method described in Section \ref{sec:results}. The 
   histogram yields $12 \lesssim K \lesssim 117$  as the
   $99\%$ credible interval for the $K$ values of possible migration scenarios, similar to the estimated bounds from 
   the eccentricity track derived using $N$-body simulation
   in Figure \ref{fig:posterior_ecc_track}.
   }
   \label{fig:K_histogram}
\end{figure*}

Finally, we also explicitly compute a corresponding $K$ value for each sample in the posterior distribution, assuming it is in an equilibrium MMR configuration. For each posterior sample, we substitute the planet masses and eccentricities into Equation \eqref{eqn:dissipative_d} and solve for the value of $K = \tau_\alpha/\tau_e$ that gives $\frac{d}{dt}\mathcal{D}\bigg|_\mathrm{dis} = 0$. 
While many posterior samples are not located exactly on the equilibrium line, the deviation is small so we nevertheless assign them the $K$ value of the associated MMR configuration.
Computing the $K$ values of all posterior samples in this manner, we obtain the distribution shown in Figure \ref{fig:K_histogram}.
The $99\%$ credible interval of $12 \lesssim K \lesssim 117$  for the  distribution of $K$ values  in Figure \ref{fig:K_histogram} is generally in agreement with the $3\sigma$ bounds estimated by comparing the posterior eccentricities to $N$-body simulations in Figure \ref{fig:posterior_ecc_track}.

\section{Discussion and  Conclusions} \label{sec:discussion_conclusions}

In this paper, we reanalyzed the recently extended data set of RV measurements from the planetary system HD 45364 and fit an RV signal to observations using an $N$-body dynamical model. 
We also investigated how the inferred dynamical state of the system can be influenced by the choice of model priors and by the assumed level of measurement noise.
Finally, we explored possible scenarios for the migration history of HD 45364's planets assuming an MMR configuration of the system.

Our conclusions are summarized as follows: 
\begin{enumerate}
    \item 
    We demonstrate that model priors have a large influence on the inferred MMR state of HD 45364 by comparing the results of an
    MCMC simulation that adopted a typical uninformative prior 
    to one that adopted a prior more strongly weighted toward small libration amplitudes
    (i.e., resonant configurations). Furthermore, failure to properly account for instrumental jitter can lead to biased inference of libration amplitudes in general.
    In the case of HD 45364 specifically, orbital configurations with small libration amplitudes become disfavoured relative to those with large libration amplitudes if instrumental jitter is not appropriately accounted for.
    More observations in the future will help conclusively determine whether the system is deep in the 3:2 MMR.
    \item 
    We find that the current orbital configuration of the HD 45364 system
    is consistent with past smooth orbital migration and eccentricity damping
    resulting in capture into MMR. 
    Furthermore, we estimate 
    the ratio of migration to eccentricity damping timescales, $K = \tau_a/\tau_e$, during this process 
    was in the range $11 \lesssim K \lesssim 144$.
    \item The large masses and closely spaced orbits of the planets in HD 45364 produce strong planet-planet gravitational interactions, which we leverage in our $N$-body dynamical model to break the $m\sin I$ degeneracy present in most RV observations.
    This allows us to constrain the masses of the planets to within a factor of $\sim$$1.5$  of their median values and the inclination of the system to $\sin I \gtrsim 0.6$.
\end{enumerate}
We emphasize that, at present, the data do not conclusively constrain the HD 45364 planets to be locked in the 3:2 MMR. Nonetheless, dynamical models with priors strongly weighted toward low libration amplitude configurations, the expected outcome of resonant capture via smooth migration, produce a quality of fit to the existing data that is nearly indistinguishable from the quality of fit obtained while adopting conventional uninformative priors. In principle, the full mathematical machinery of Bayesian model comparison could be deployed to ascribe a likelihood that the system's planets reside in resonance. However, any quantitative determination of this likelihood would require apportioning prior probabilities to competing models with different assumptions about the system's formation history. Since the current state of planet formation theory is far too uncertain to yield robust predictions for such prior probabilities, we have forgone any such quantitative Bayesian model comparison calculations. Instead, we simply note that the RV data for HD 45364 are wholly consistent with theoretical expectations for resonant planetary systems formed through a process of smooth, disk-induced migration.
We envisage future application of the methods developed in this paper to other candidate systems with  planets in or near MMR, including RV-detected systems \citep[e.g.][]{wittenmyer_pan-pacific_2016, luque_precise_2019, rosenthal_measuring_2019, trifonov_two_2019}, transiting planet pairs \citep[e.g.][]{dawson_precise_2021, bozhilov_21_2023} and higher-multiplicity resonant chains \citep[e.g.][]{steffen_transit_2013, mills_resonant_2016, luger_seven-planet_2017, leleu_six_2021, macdonald_five-planet_2021, dai_toi-1136_2023}
in order to better understand the role of planetary migration in the formation history of these systems.
In addition, we hope that our findings about the relationship between unmodeled RV measurement noise (instrumental jitter) and inferred libration amplitude will help inform studies of other resonant planetary systems.

\section{Acknowledgments} 
\label{sec:acknowledgments}
We thank Hanno Rein for helpful discussions. We thank Trifon Trifonov for a careful referee report that helped improve this manuscript.
S.H. acknowledges support by the Natural Sciences and Engineering Research Council of Canada (NSERC), funding references CITA 490888-16 and RGPIN-2020-03885.

\software{
\texttt{REBOUND} \citep{rein_rebound_2012}, 
\texttt{emcee} \citep{foreman-mackey_emcee_2013},
\texttt{REBOUNDx} \citep{tamayo_reboundx_2020}
}
 
\bibliography{HD45364_paper}{}
\bibliographystyle{aasjournal}

\end{document}